\documentclass[lettersize,journal]{IEEEtran}
\usepackage{amsmath,amsfonts}
\usepackage{algorithmic}
\usepackage{algorithm}
\usepackage{array}
\usepackage{cite}
\usepackage[caption=false,font=normalsize,labelfont=sf,textfont=sf]{subfig}
\usepackage{textcomp}
\usepackage{stfloats}
\usepackage{url}
\usepackage{bm}
\usepackage{verbatim}
\usepackage{graphicx}
\usepackage{cite}
\usepackage{fancyhdr}
\usepackage{multirow} 
\usepackage{makecell}
\usepackage{booktabs}    
\usepackage{titlesec}
\usepackage{times}
\usepackage{wasysym}
\usepackage{color}
\usepackage{colortbl}
\newtheorem{remark}{Remark}

\begin{document}
\title{Enhanced Koopman Operator Approximation for Nonlinear Systems Using Broading Learning System}

\author{YANGJUN SUN, ZHILIANG LIU \thanks{YANGJUN SUN and ZHILIANG LIU are with the Institute of Complexity Science and the School of Automation, Qingdao University, Qingdao 266071, China (e-mail: sunyangjun@qdu.edu.cn; liuzhiliang@qdu.edu.cn).}}



\maketitle

\begin{abstract}
Traditional control methods often show limitations in dealing with complex nonlinear systems, especially when it is difficult to accurately obtain the exact system model, and the control accuracy and stability are difficult to guarantee. To solve this problem, the Koopman operator theory provides an effective method to linearise nonlinear systems, which simplifies the analysis and control of the system by mapping the nonlinear dynamics into a high-dimensional space. However, the existing extended dynamical mode decomposition (EDMD) methods suffer from randomness in the selection of basis functions, which leads to bias in the finite-dimensional approximation to the Koopman operator, thus affecting the accuracy of model prediction. To solve this problem, this paper proposes a BLS-EDMD method based on the Broad learning system (BLS) network. The method achieves a high-precision approximation to the Koopman operator by learning more accurate basis functions, which significantly improves the prediction ability of the model. Building on this, we further develop a model predictive controller (MPC) called BE-MPC. This controller directly utilises the high-dimensional and high-precision predictors generated by BLS-EDMD to predict the system state more accurately, thus achieving precise control of the underwater unmanned vehicle (UUV), and its effectiveness is verified by simulation. 
\end{abstract}


\section{Introduction}

Modeling and control of complex nonlinear systems is a long-standing challenge in various fields \cite{electronics9060940,wang2022adaptive,meng2023epch,bwambale2023data}. Traditional methods, such as PID \cite{coskun2023intelligent}, LQR \cite{muhssin2023optimal}, adaptive control \cite{wang2023adaptive}, and MPC \cite{lawrynczuk2024koopman,wei2022mpc,nguyen2024efficient} are very effective in relatively simple environments \cite{xu2021underwater,li2024path}, but struggles when dealing with nonlinear and time-varying conditions, especially in systems where accurate mathematical models are difficult to obtain \cite{yuan2022event}. To address these challenges, the Koopman operator-theoretic approach \cite{koopman1931hamiltonian} has emerged as a strong alternative. Unlike traditional control strategies that require detailed system models, Koopman theory provides a framework for linearizing nonlinear dynamics by mapping them into a high-dimensional space.

Koopman operator theory can simplify the modeling of nonlinear systems more effectively compared to traditional methods. It maps nonlinear dynamics to a linear space, achieving linearization of the system's behavior, which is inherently linear and infinite-dimensional in nature. This theory has been applied to various fields such as fluid dynamics \cite{WOS:000520951900011,li2023koopman}, grid analysis \cite{gong2023novel,nandanoori2022graph}, and biomedical engineering \cite{liang2022online,golany202112}, among others. However, practical applications require finite-dimensional representations. Dynamic Mode Decomposition (DMD) \cite{snyder2021koopman}, which uses Singular Value Decomposition (SVD) to identify the main dynamic modes in the data, corresponds to the key dynamic behavior of the system. Yet, because it is based on the assumption of linear mapping, DMD struggles to fully capture the characteristics of a nonlinear system. In the field of dynamics, EDMD \cite{williams2015data} has proven effective in extracting key features and modes from data of complex dynamic systems, offering good stability and controllability \cite{yang2024model}. The idea is to select a set of basis functions to lift the system to a higher-dimensional space, thereby linearizing it, and then use these functions to approximate the Koopman operator. In recent years, the extended Koopman operator method \cite{proctor2018generalizing,guo2023koopman} based on EDMD has been utilized to construct high-dimensional predictors and has been combined with MPC for system simulation and control \cite{toro2023data,vsvec2021model,shi2023koopman,shi2021enhancement}. However, in the absence of prior knowledge, the selection of basis functions may lead to inaccurate finite-dimensional approximations of the Koopman operator, introducing prediction errors that can subsequently affect the performance of the MPC.

\begin{figure*}[htbp] 
	\centering 
	\includegraphics[width=14cm]{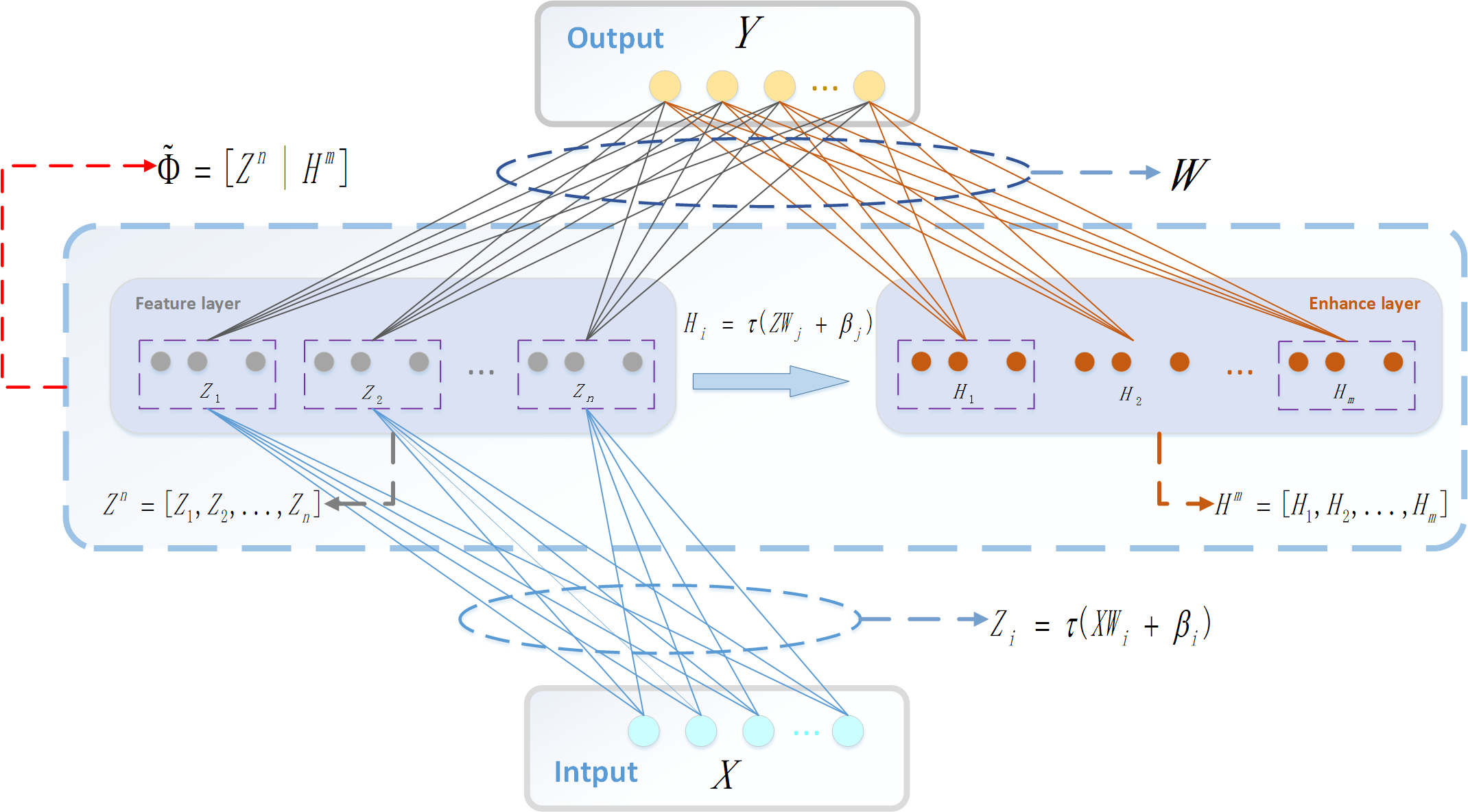} 
	\caption{The hidden layer $\varphi$ of the Broad Learning System network consists of the feature layer $Z^n$ and the enhancement layer $H^n$. The input $x$ is enriched by $\varphi$ to better represent the feature.} 
	\label{BLS} 
\end{figure*}

Recently, Artificial Neural Networks (ANNs) have successfully addressed the challenge of learning basis functions from data. A common method involves constructing an autoencoder \cite{lusch2018deep} to represent both the basis function and its inverse \cite{heijden2020deepkoco,otto2019linearly}. This approach employs a multilayer feedforward network to generate basis functions, as detailed in \cite{yeung2019learning}. The generated basis functions mitigate the influence of randomness, achieved through the structure of the multilayer feedforward networks. In the paper \cite{han2020deep}, deep neural networks (DNNs) are used to learn these basis functions, and the effectiveness of this method is validated. However, constructing nonlinear models often requires a large amount of snapshot data, resulting in multilayer deep networks that can take dozens of hours to update. The Broad Learning System (BLS) network proposed by \cite{chen2017broad} offers a simpler structure compared to deep networks, featuring only one hidden layer, which consists of a feature layer and an enhancement layer. This structure reduces the number of network weight updates, improving efficiency while retaining all the characteristics of the nonlinear system \cite{liu2021incremental,ZHANG2024102522}. The BLS network structure includes an input layer, a hidden layer $\Phi$, and an output layer, as shown in Fig. \ref{BLS}. A distinctive feature of this network structure is that $\Phi$ is further subdivided into a feature layer $Z^n$ and an enhancement layer $H^m$, which are key regions where nonlinear operations are introduced to enhance the model's expressive power \cite{WOS:000524916400001}. The BLS network represents an innovative improvement over the Random Vector Functional Linkage Network (RVFLNN) \cite{shi2021random}. In the data processing flow, the input data first undergoes a nonlinear feature extraction process $ Z_{i}=\tau\left(X W_{e i}+\beta_{e i}\right) $, which aims to extract more abstract and discriminative features from the raw data to form the feature layer $Z^n$. Subsequently, the feature layer $Z^n$ generates a set of $m$ augmentation nodes $H^m$ through a nonlinear augmentation mapping mechanism $H_{j}=\tau\left(Z^{n} W_{h j}+\beta_{h j}\right) $. Ultimately, the BLS uses pseudo-inverse computation of the hidden layer and the output to derive the output weights $W=\Phi ^\dagger Y$, enriching the models feature representation and enhancing the networks ability to capture complex patterns. Compared to deep neural networks, the structural design of the BLS network is simpler and more intuitive. It avoids the complex multilayer structure and the intricate backpropagation process common in deep networks \cite{WOS:000780850000003}.

To address the aforementioned problem, this paper proposes a BLS-based EDMD (BLS-EDMD) approach to approximate the Koopman operator. The algorithm utilizes two BLS: one for encoding during training and another for decoding during prediction. Unlike traditional methods, the number of hidden nodes in the BLS Networks can be flexibly adjusted according to the complexity of the training task without compromising the stability and expressiveness of the overall network structure. Furthermore, the BLS-EDMD method improves adaptability to nonlinear systems by introducing a set of generalized basis functions, which helps reduce the finite-dimensional approximation error caused by improper basis function selection. In this paper, the high-dimensional predictor obtained using the BLS-EDMD method is further employed to design a novel MPC for target tracking control of a Deep Sea Rescue Vehicle (DSRV). This MPC framework fully leverages the system state information provided by the high-dimensional predictor to achieve more accurate control strategies, and its performance is validated through MATLAB simulations.

The contributions of this article are summarized as follows.

1) This method leverages the BLS to configure hidden layers, avoiding errors in basis function selection and improving system state representation by adjusting the number of hidden nodes without altering the network structure.

2) Based on the proposed method, a high-dimensional linear predictive model is obtained, which is then used to design an MPC control algorithm, and its effectiveness has been verified through UUV tracking control simulations.

3) The proposed method is simple and efficient, ensuring the accuracy of predictions without the need for lengthy training processes, offering a new perspective for the use of algorithms similar to EDMD.

\section{Modeling methods for BLS-EDMD}
In this section, we present a methodology for utilizing EDMD to represent the Koopman operator. Furthermore, we delineate the structure of a high-dimensional predictor and introduce the feature extraction capabilities of the BLS network. This includes its feature and enhancement layers, which facilitate the generation of rich feature-learning basis functions.

\subsection{EDMD approximation of the Koopman operator}
The discrete form of a general nonlinear system is given as follows:
\begin{align}
	x_{k+1}=f(x_k,u)
	\label{1}
\end{align}
where $x_k \in \mathbb {R} ^ n $ is the state vector of the system, $u \in \mathbb {R} ^ l $ is the control input vector of the system, and $f: \mathbb {R} ^ n \times \mathbb {R} ^ l \to \mathbb {R} ^ n $ describes the system's evolution. The presence of control inputs complicates the system's dynamics, necessitating extensions to the theory of Koopman operators. This paper offers feasible solutions by combining Koopman operators with MPC, and defines an extended state :

\begin{align}
	\xi_k =\begin{bmatrix}
		x_k\\
		u_k
	\end{bmatrix}
	\label{2}
\end{align}

The system (\ref{1}) can be rewritten as follows:
\begin{align}
	\xi_{k+1}=f(\xi_k)
	\label{3}
\end{align}
where $\xi_k$ is the extended state at step k, adapted here to the Koopman theory of autonomous systems. Based system (\ref{3}) that collects snapshot data:
\begin{align}
	X&=[x_1,x_2,\dots ,x_N]\\
	Y &=[y_1,y_2,\dots ,y_N]\\
	U &=[u_1,u_2,\dots ,u_N]\\
	\mathbf{\chi_x}& =[\xi_1,\xi_2,\dots ,\xi_N]
	\label{4}
\end{align}
where $Y$ is the next moment state of $X$, $\chi_x$ denotes the extended state, and $N$ is the number of supersamples. Continue by defining a set of basis functions to represent the linear representation of the Koopman operator in this system:

\begin{align}
	\Phi(x_k) =[{\varphi_1(x_k)}^T,{\varphi_2(x_k)}^T,\dots ,{\varphi_M(x_k)}^T]^T
	\label{5}
\end{align}
where $\Phi(x_k)\in \mathbb {R} ^{M}$, $M$ is the number of observable functions on $x$, the action of the Koopman operator on these basis functions can be expressed as follows:

\begin{align}
	\mathcal{K} \Phi(\xi_{k})=\Phi(\xi_{k+1})=\Phi(f(\xi_{k}))
	\label{6}
\end{align}

The predictor under Koopman theory is defined as:

\begin{align}
	\left\{\begin{matrix}
		\Phi (x_{k+1})=A\Phi (x_{k})+Bu_k \\
		\hat{x} _k=C\Phi (x_{k})
	\end{matrix}\right.
	\label{8}
\end{align}
where  $A \in \mathbb {R} ^{ M\times M}$, $B  \in \mathbb {R} ^{ M\times l}$ and $C \in \mathbb {R} ^{ n\times M}$, $\hat{x}_k$ is the expected value of $x_k$ mapped and returned in the Koopman operator.

Since the $A$ and $B$ in the prediction are highly correlated with the infinite-dimensional matrix $\mathcal{K}$, we can approximate the action of the Koopman operator $\mathcal{K}$ with a finite-dimensional matrix $\tilde{\mathcal{K}} $. To do this, we define the following matrix:

\begin{align}
	\Phi(\chi_x)^T = \begin{bmatrix}
		\varphi_1(\xi_1) & \varphi_2(\xi_1) & \cdots & \varphi_M(\xi_1) \\
		\varphi_1(\xi_2) & \varphi_2(\xi_2) & \cdots & \varphi_M(\xi_2) \\
		\vdots & \vdots & \ddots & \vdots \\
		\varphi_1(\xi_N) & \varphi_2(\xi_N) & \cdots & \varphi_M(\xi_N)
	\end{bmatrix}
	\label{7}
\end{align}

Introducing the L2 regularization term to construct the optimization problem under the EDMD method , we approximate $\mathcal{K}$ as:

\begin{align}
	\min_{\tilde{\mathcal K} }\sum_{k=1}^{N} \left \|  \Phi (\xi_{k+1})-\tilde{\mathcal K}\Phi (\xi_k) \right \|_2^2
	\label{9}
\end{align}
where $\tilde{\mathcal{K}}\in \mathbb {R} ^ S$ is an approximation of the Koopman operator, $S=M+l$. In addition, we would like to obtain the prediction results under Koopman's theoretical study, (\ref{9}) variant as:
\begin{align}
	\min_{A,B} \sum_{k=1}^{N} &\left \| \begin{bmatrix}
		\Phi(x_{k+1}) \\ \nonumber
		u_{k}
	\end{bmatrix} -\begin{bmatrix}
		A& B\\
		\dots  &\dots 
	\end{bmatrix}\begin{bmatrix}
		\Phi(x_{k}) \\
		u_{k}
	\end{bmatrix}\right \|_2^2 \\
	\label{10}
\end{align}
where we disregard the variation of the basis function $\Phi$ with respect to $u$ such that $\tilde{\mathcal{K}} =\begin{bmatrix}
	A& B\\\\
	0 &I
\end{bmatrix}$, By solving (\ref{10}) optimization problem, we can find the $A$ and $B$ matrices which approximate the Koopman operator for a controlled nonlinear system.

Let $\Phi_U=\begin{bmatrix}
	\Phi(X)\\
	U
\end{bmatrix}$ and $\Psi _U=\begin{bmatrix}
	\Phi(Y)\\
	U
\end{bmatrix}$ be the concatenated matrix of the basis functions and control inputs. The optimization problem then becomes:
\begin{align}
	\min_{A,B}  &\left \| \Phi(Y)  -\begin{bmatrix}
		A& B
	\end{bmatrix}\Phi_U(X)\right \|_F
	\label{11}
\end{align}

By solving (\ref{11}), the analytic solutions for $A$ and $B$ are
\begin{align}
	\begin{bmatrix}
		A&B
	\end{bmatrix}=\Phi(Y)\begin{bmatrix}
		\Phi(X) ,U
	\end{bmatrix}^\dagger 
	\label{12}
\end{align}

We wish to find the matrix $C$, introduce the L2 regularization term and to construct the optimization problem:
\begin{align}
	\min _{{C}}\left\|{X}-{C} \Phi(X) \right\|_{F}^{2}
	\label{13}
\end{align}
The analytic solution of the optimization problem (\ref{13}) is:
\begin{align}
	C = X \Phi(X)^\dagger 
	\label{14}
\end{align}

\subsection{BLS-EDMD learns the Koopman operator}

The selection of basis functions in EDMD for approximating Koopman theory is typically done empirically. Common basis functions include radial basis functions, polynomial basis functions, Gaussian basis functions, and others. However, different basis functions can have varying impacts on the extraction of system features. In this section, we will introduce the BLS-EDMD algorithm, which learns the basis functions to enhance the performance of the EDMD approximation of Koopman theory.
\begin{figure*}[htbp] 
	\centering 
	\includegraphics[width=12cm]{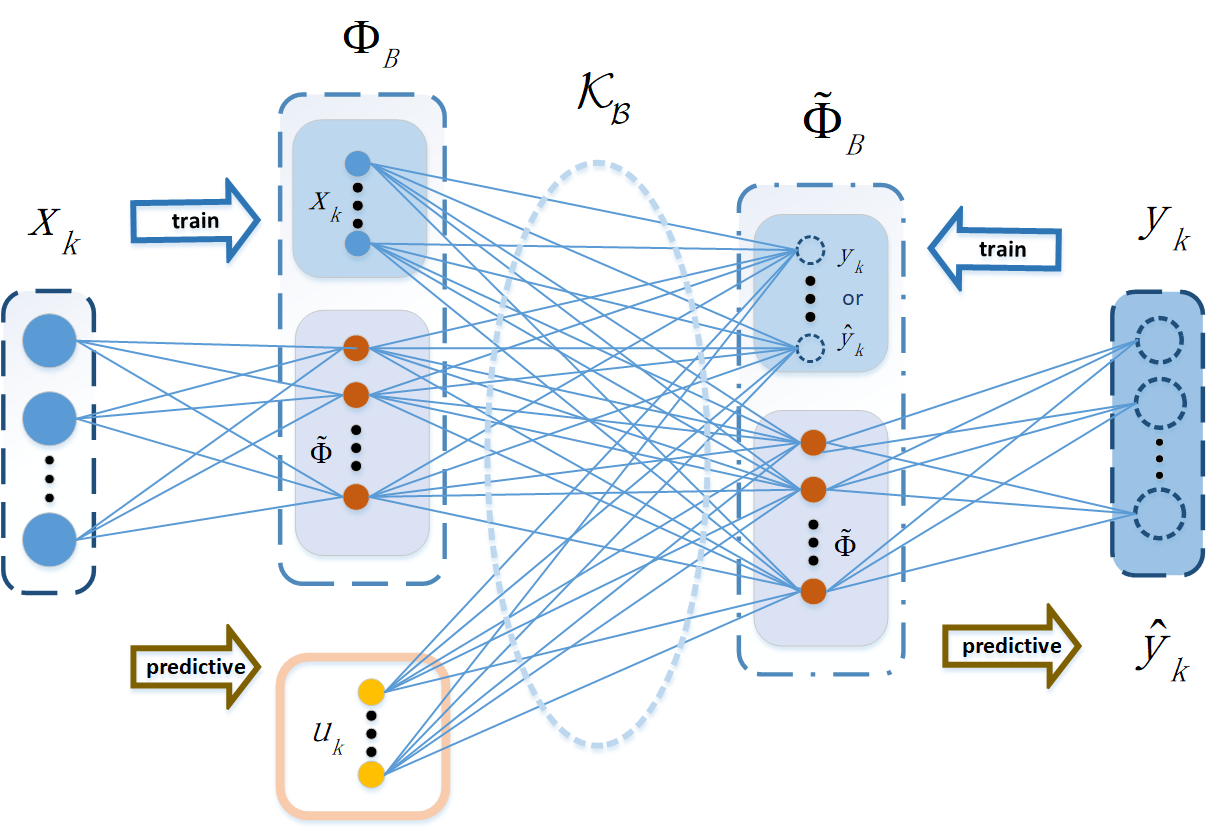} 
	\caption{The BLS-EDMD method plays a role in lift the dimensionality of both $\Phi_B $ and $\tilde\Phi_B $ during the training process. In the training part, the ridge regression operation is performed by boosting states $x$ and $y$ to obtain $\mathcal{K_B}$,In the prediction part, $x_k$ lift states $\begin{bmatrix}
			x_k \\
			\tilde\Phi(x_k)
		\end{bmatrix}$ is mapped by $\mathcal{K_B} $ to $\begin{bmatrix}
			\hat{y}_k \\
			\tilde\Phi(y_k)\end{bmatrix}$ and decoded by $\tilde\Phi_B$ to get $\hat y_k$.} 
	\label{????} 
\end{figure*}

The BLS-EDMD method contains both training and prediction components, as shown in Fig. \ref{????}. In the training process, the input of the system is the state data $x_k$ , firstly, the input $x_k$ passes through the hidden layer of Fig. \ref{BLS} to generate the feature mapping $\tilde\Phi(x_k)$ and combines with $x_k$ to form a new feature mapping $\Phi_B(x_k)$.
\begin{align}
	\Phi_B(x_k)=\begin{bmatrix}
		x_k \\\
		\tilde\Phi(x_k)
	\end{bmatrix}
	\label{x}
\end{align}

Similarly, the target output matrix $y_k$ is passed through the hidden layer of Fig. 1 to generate the feature mapping $\tilde\Phi(y_k)$, which is combined with $y_k$ to generate a new feature mapping $\tilde\Phi_B(y_k)$:
\begin{align} \tilde\Phi_B(y_k)=\begin{bmatrix}
		y_k \\\
		\tilde\Phi(y_k)
	\end{bmatrix}
	\label{y}
\end{align}
Next, the Koopman operator $\mathcal {K_B}$ is solved using the following optimization objective:
\begin{align}
	\min_{\mathcal {K_B} } \left \| \mathcal {K_B}\xi(x_k)-\tilde{\Phi} _B(y_k)\right \|^2_2 
	\label{K_B}
\end{align}
where $\xi(x_k)=\begin{bmatrix}
	\Phi_B(x_k)\\\\
	u_k
\end{bmatrix} $, $u_k$ is the control input to the system and the objective is to minimize the two-paradigm error between the output after the action of $\mathcal{K_B}$ and the target eigenvector $\tilde\Phi_B(y_k)$. Through this optimization process, the Koopman operator $\mathcal{K_B}$ can be obtained for subsequent prediction tasks.

In the prediction stage, the new system state is predicted using the Koopman operator $\mathcal{K_{B}}$ obtained in the training stage. Firstly, the new state data $x_{k}$ is inputted to the BLS network to generate the feature mapping $\tilde\Phi(x_{k})$ and combined with $ x_{k} $ to form $\Phi_{B}(x_{k}) $ . Subsequently, this feature vector is updated using $ \mathcal{K_{B}} $ to obtain the predicted augmented feature vector $ {\Phi}_{B}(\hat{y}_{k}) $ :
\begin{align}
	\tilde \Phi_B(\hat{y}_k)=\mathcal{K_B}\xi(x_k)
\end{align}

Finally the predicted state of the system $\tilde{\Phi}_{B}(\hat y_{k}) $ is extracted from the augmented feature vector $\hat{y}_{k} $ , which provides the necessary basis for the decision making of the subsequent MPC.

In Fig. \ref {????}, the basis function is defined as

\begin{align}
	\Phi_B =[\tilde{\varphi}_1^T,\tilde{\varphi} _2^T,\dots ,\tilde{\varphi} _{E}^T]^T
	\label{15}
\end{align}
where the $E$ size depends on the number of network features $Z$ and augmentation nodes $H$ points, and based on the theoretical work of BLS-EDMD Koopman, the high-dimensional predictors (\ref{8}) are redefined as:
\begin{align}
	\left\{\begin{matrix}
		\Phi_B(y_{k})=\mathcal{K_B}\xi(x_k)\\
		\hat{y}_{k} =C_b{\Phi }_B(y_k)
	\end{matrix}\right. 
	\label{16}
\end{align}
where $ \mathcal{K_B}\in\mathbb{ R} ^{E \times P}$, $C_b\in \mathbb{R} ^{n\times E}$ and $P=E+l$. Let $ \mathcal{K_B}=[A_b\ \ \ B_b]$ , where $A_b\in \mathbb{R} ^{E\times E}$, $B_b\in \mathbb{R} ^{E\times l}$ and furthermore get the high-dimensional predictor by (\ref{16}).
\begin{align}
	\left\{\begin{matrix}
		\Phi_B(y_k)=A_b\Phi_B(x_k)+B_bu_k\\
		\hat y_{k}=C_b\Phi_B (y_k)
	\end{matrix}\right. 
	\label{yuce}
\end{align}

According to the EDMD theory we can get the expressions for $[A_b \ B_b]$ and $C_b$ in BLS-EDMD.
\begin{align}
	\min_{A_b,B_b}\left \| \Phi_B(Y)-A_b\Phi_B(X)-B_bU \right \|_F 
	\label{A}
\end{align}
\begin{align}
	\begin{bmatrix}
		A_b&B_b
	\end{bmatrix}=\Phi_B(Y)\begin{bmatrix}
		\Phi_B(X),U
	\end{bmatrix}^\dagger 
	\label{AB}
\end{align}
\begin{align}
	\min_{C_b}\left \| \Phi_B(X)-C_b\Phi_B(X) \right \|_F
	\label{C}
\end{align}
\begin{align}
	C_b = X \Phi_B (X)^\dagger 
	\label{Cb}
\end{align}

\begin{remark}
	Note that the lifting function $\Phi_B$ includes observable state quantities, then the solution of (\ref{Cb}) can be simply expressed as $C = [I, 0]$, where $I$ is the identity matrix of size $n$. This means that $C$ only needs to extract the original state variables without dealing with any additional lifted variables.
\end{remark}	


\begin{algorithm} 
	\caption{Training Steps for BLS-EDMD Network} 
	\label{alg:broad_edmd} 
	\begin{algorithmic}
		\REQUIRE Initialise $A_b$, $B_b$ and the number of nodes $N_{init}$ (including $Z^n$ and $H^m$ nodes), set the error threshold $\epsilon > 0$ and the $error = \infty$
		\ENSURE Trained $A_b$, $B_b$, $C_b$
		\STATE Sample the state and inputs, i.e., $X = [{x_{1:m}^{[j]}}]_{j=1}^N$, $Y = [{y_{1:m}^{[j]}}]_{j=1}^N$, $U = [{u_{1:m}^{[j]}}]_{j=1}^N$. 
		\WHILE{{$error \geq \epsilon$}} 
		\STATE  Obtain ascending state $\Phi_B(x_k)$ for $x_k$ using (\ref{x}) and $\Phi_B(y_k)$ for $y_k$ using (\ref{y}) 
		\STATE Compute $\mathcal{K_B}$ using (\ref{K_B}) 
		\STATE Calculate $\hat{y}_k$ using (\ref{yuce}) 
		\STATE Calculate the training $error = \left\| \hat{y}_k - y_k \right\|^2$ 
		\IF{$error \geq \epsilon$} 
		\STATE Increase $Z^n$ and $H^m$ node counts  
		\STATE Reset $error = \infty$ 
		\ENDIF 
		\ENDWHILE 
	\end{algorithmic} 
\end{algorithm}

\section{Predictive control }
Combined with the high-dimensional predictor designed using the BLS-EDMD method, we can more accurately describe the dynamic behavior of the nonlinear system and provide high-precision state predictions for the  MPC. In this section, we will describe the design process of the BE-MPC controller in detail.

In this section we propose BE-MPC controller based on BLS-EDMD and we set the objective function based on (\ref{yuce}) predictor
\begin{align}
	\min_{u} J=\sum_{i=1}^{N_t} (y_i-y_{ref,i})^TQ(y_i-y_{ref,i})+u_i^TRu_i
	\label{youhua}
\end{align}
where the sliding window size is set to $N_t$, $y_{ref,i}$ denotes the target value at the $i$th step, and $Q\in \mathbb{R} ^{n\times n}$, $R\in \mathbb{R} ^{m\times m}$ denote the positive definite matrices for the prediction output error and the control input, respectively. In the optimisation we consider the following constraints:
\begin{align}
	y_{min}<y_i<y_{max}
\end{align}
\begin{align}
	u_{min}<u_i<u_{max}
\end{align}

The prediction based on the output of the high dimensional predictor (\ref{yuce}) can be expressed as
\begin{align}
	Y_i=\Upsilon \tilde{\varphi } (x_i)+\Omega U_i 
	\label{y_i}
\end{align}
where \\
\begin{align}
	Y_i=\begin{bmatrix}
		y_1^T&  y_2^T&\cdots   &y_{N_t}^T
	\end{bmatrix}^T
\end{align} 
\begin{align}
	\Upsilon=\begin{bmatrix}
		C_bA_b &  C_bA_b^2 & \cdots & C_bA_b^{N_t}
	\end{bmatrix}^T
\end{align}
\begin{align}
	\Omega =\begin{bmatrix}
		C_bB_b & 0 & \cdots   & 0\\
		C_bA_bB_b& C_bB_b &\cdots   & 0\\
		\vdots & \vdots  &\ddots   & \vdots \\
		C_bA_b^{N_t-1}B_b& C_bA_b^{N_t-2}B_b & \cdots  &C_bB_b
	\end{bmatrix}
\end{align}
To facilitate optimisation, the (\ref{youhua}) optimisation problem can be expressed as:
\begin{align}
	\min_{U_k}\frac{1}{2} U_k^TSU_k +G_tU_k
	\label{U}
\end{align}
where the target sequence $Y_{ref,i}=\begin{bmatrix}
	y_{ref,1}&  y_{ref,2}& \cdots  &y_{ref,N_t}
\end{bmatrix}$, $\mathcal{E}_t =\Upsilon \tilde{\varphi } (x_k)-Y_{ref,i}$ and $G_t=2\mathcal{E}_t^TQ\Omega$, $S=\Omega ^TQ\Omega +R$.

\begin{algorithm}
	\caption{BE-MPC Control Process}
	\label{alg:broad_edm}
	\begin{algorithmic}
		\REQUIRE Trained matrices $A_b$, $B_b$; control window size $N_t$; weights $Q$, $R$; initial state $X_0$;
		\ENSURE Optimal control sequence $U$;
		\FOR {$k = 0$ to $N-1$}
		\STATE Compute the lifted state $\Phi_B(X_k)$ and set the reference trajectory $Y_{ref}$;
		\STATE Calculate the predicted output $Y_i$ for the window size $N_t$ using the (\ref{yuce});
		\STATE Solve the optimization problem with constraints (\ref{U}) to get the optimal control sequence $U_k$;
		\STATE Apply the first control input $U_k(0)$ to the nonlinear system;
		\STATE Update current status to $X_{k+1}$;
		\ENDFOR
	\end{algorithmic}
\end{algorithm}

\begin{figure}[!t]  
	\centering  
	\subfloat{\includegraphics[width=3.1in]{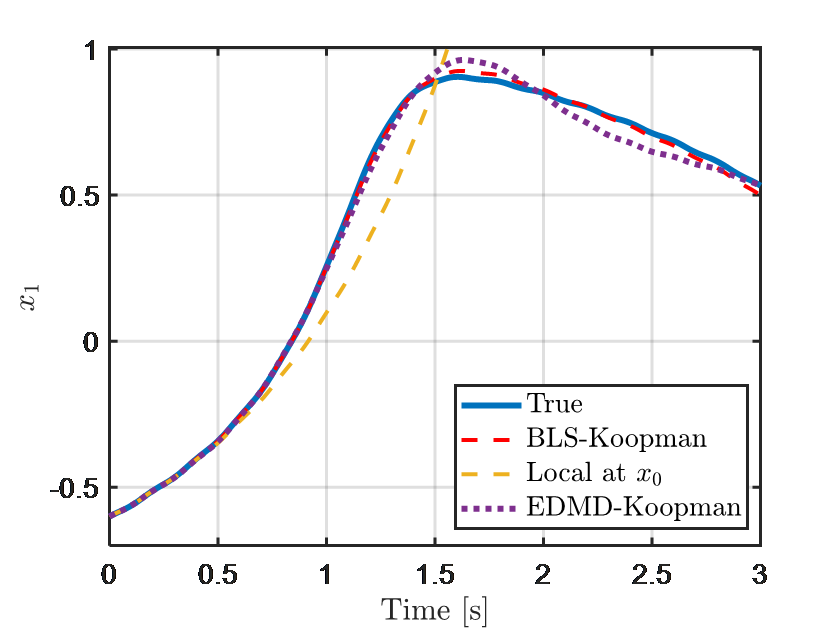}
		\label{fig_1a}}
	\hfill  
	\subfloat{\includegraphics[width=3.1in]{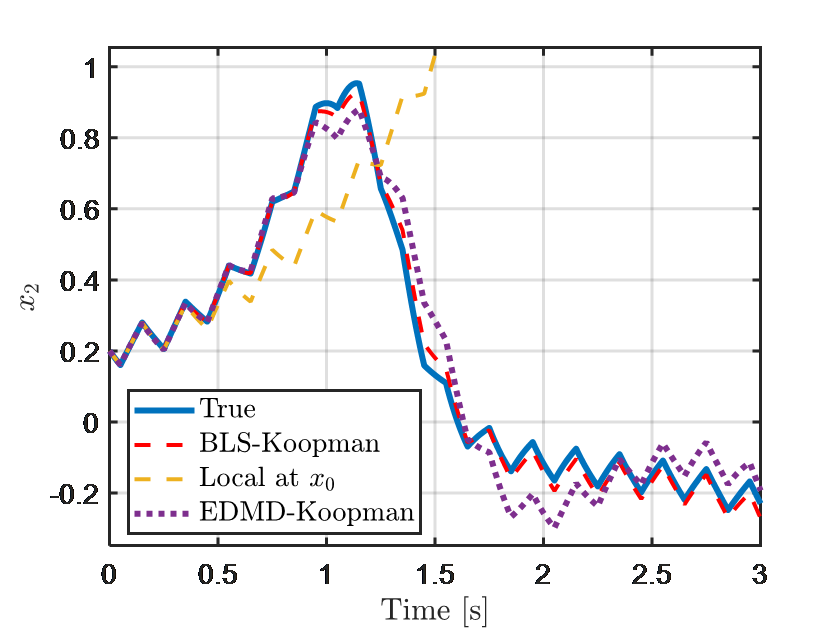}
		\label{fig_2b}}
	
	\caption{Prediction comparison based on forced van der Pol oscillator: The initial condition $x_0 =[-0.6,0.2] ^T$ set in the interval [-1,1] and a simulation time of $3$s, we performed a prediction comparison. In this process, the control input u(t) is set to be a square wave signal with a period of 0.3 s and unit amplitude. }  
	\label{BE-K}  
\end{figure}  
\section{Simulations}

In this section, the proposed high-dimensional predictor (\ref{15}) method (i.e., BLS-Koopman) is validated in the matlab environment. The performance is also compared with the EDMD-based Koopman method (EDMD-Koopman) and the predictor based on dynamic local linearization for a given initial condition $x_0$. The UUV target tracking control is implemented based on the controller proposed in Section III.
\subsection{Prediction}
For the comparison of the prediction part we choose classical forced van der Pol vibronic system:
\begin{align}
	\dot x_1&=-2x_2 \nonumber \\ 
	\dot x_2&=0.8x_1 + 10x^2_1x_2 - 2x_2 + u 
\end{align}
\begin{figure}[!t]  
	\centering  
	\subfloat{\includegraphics[width=3.1in]{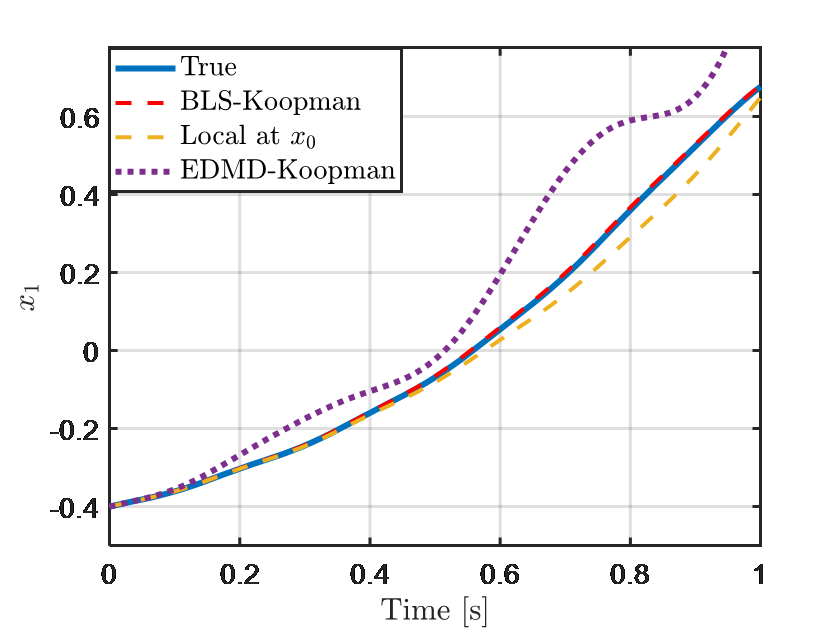}
		\label{fig1a}}
	\hfill  
	\subfloat{\includegraphics[width=3.1in]{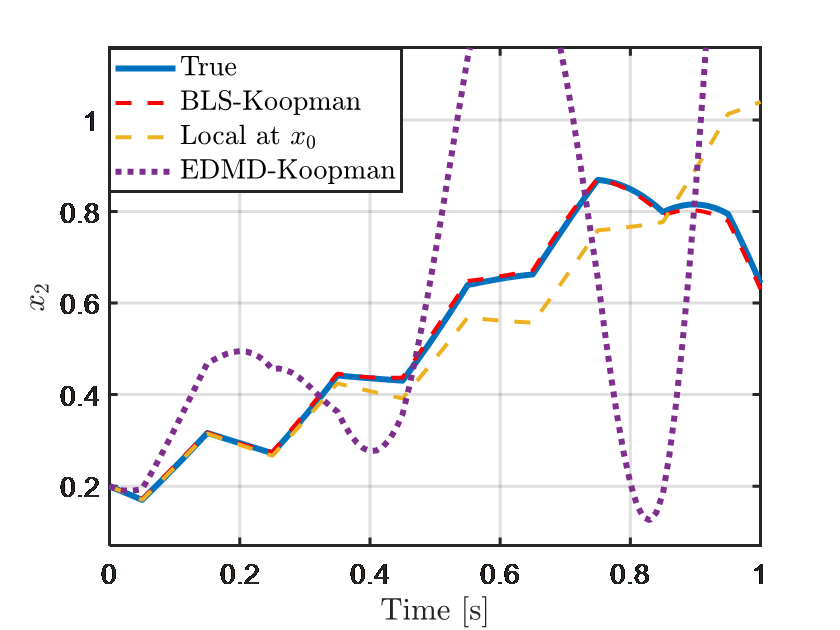}
		\label{fig2b}}
	
	
	\caption{Robustness test based on forced van der Pol oscillator: The initial condition ${x_0} = [-0.4, 0.2]^T$ is set in the interval $[-0.5,0.5]$, and the simulation time is $1$s. We perform a robustness comparison. In this process, the control input u(t) is set to be a square wave signal with a period of 0.3 s and unit amplitude. }  
	\label{lubang}  
\end{figure}

During network prediction, we collect 300 simulation steps with 500 trajectories per step. The control inputs for each trajectory are $[-1,1]$ random numbers, the number of $\tilde\Phi$ nodes is 1000 (where the number of feature nodes and enhancement nodes are 600 and 400, respectively), and the matrices $\tilde\Phi(x_k)$ and $\Phi_B(x_k)$  of sizes $1000$ and $1002$. For comparison, we choose the Thin plate spline radial basis function as the basis function of the EDMD and the activation function of the BLS network. As shown in Fig. \ref{BE-K}, for the initial state vector $x_ 0=[-0.6,0.2] ^T$ generated randomly in the interval $[-1,1]$ , comparing its true state with the predicted state obtained via the BLS-Koopman method, the local linearisation method, and the EDMD-Koopman method.

Fig. \ref{BE-K} and Fig. \ref{lubang} compare the prediction effectiveness of the BLS-Koopman and EDMD-Koopman algorithms under different initial state ranges ($[-1,1]$ vs. $[-0.5,0.5]$). The results show that the prediction performance of the EDMD-Koopman algorithm decreases significantly as the initial state changes, in contrast to the BLS-Koopman algorithm, which exhibits stronger robustness. To further validate this conclusion, for each of the three ranges $[-1,1]$, $[-0.5,0.5]$ and $[-0.8,0.8]$, $50$ different initial values are randomly selected for simulation. For each range, the average RMSE of the $50$ simulations is calculated using (\ref{RMSE}). As shown in Table \ref{rmse}, the average RMSE of the BLS-Koopman algorithm is significantly lower than that of the EDMD-Koopman algorithm, which confirms its superior stability and robustness under a broader range of superior stability and prediction accuracy under initial conditions.
\begin{align}
	\mathrm{RMSE}=\frac{1}{50}\sum_{i=1}^{50} \sqrt{\frac{\left \| \hat y_i-y_i \right \|^2_2 }{\left \| y_i \right \|^2_2 } } 
	\label{RMSE} 
\end{align}
where $\hat y_i$, $y_i$ denote the sequence of predicted values and the sequence of true values at the $i$th initial value, respectively, and $\left \| \cdot \right \| _2$ denotes the Euclidean parameter of the vector.

\begin{table}[!t]  
	\caption{Comparison of Average RMSE (\%) of Different Prediction Methods across Different State Ranges}  
	\label{tfae}  
	\centering  
	\fontsize{9pt}{12pt}\selectfont  
	\renewcommand\tabcolsep{8pt} 
	\begin{tabular}{cccc}  
		\hline  
		\textbf{Method} & \textbf{[-1, 1]} & \textbf{[-0.5, 0.5]} & \textbf{[-0.8, 0.8]} \\   
		\hline  
		BLS-Koopman    & 13.25 & 27.01 & 26.90 \\  
		Local at $x_0$ & 174.57 & 349.15 & 349.15 \\  
		EDMD-Koopman   & 15.05 & 38.93 & 32.82 \\  
		\hline  
	\end{tabular}  
\end{table}

\subsection{MPC control of DSRV}
In this section the proposed BE-MPC method is applied to the DSRV \cite{fossen2011handbook} control, and we firstly analyse its simplified model as follows:
\begin{align}
	\dot{\eta} &= J_k(\eta)\nu \nonumber \\
	M\dot{\nu} + C(\nu) + g(\eta) + g_0 &= \tau + \tau_{\text{wind}} + \tau_{\text{wave}}
\end{align}
where $\eta = [x, y, z, \phi, \theta, \psi]^T$ and $\nu = [u, v, w, p, q, r]^T$ stand for the description of the 6 degrees of freedom respectively. We are concerned primarily with the vertical motion of the DSRV, as described in \cite{fossen2011handbook}. The longitudinal subsystem can be outlined as:
\begin{align}
	\begin{bmatrix}
		\dot w\\
		\dot q\\
		\dot x\\
		\dot z\\
		\dot \theta 
	\end{bmatrix}=\begin{bmatrix}
		\frac{m_{22} Z-m_{12} M}{\operatorname{det} M}\\
		\frac{-m_{21} Z+m_{11} M}{\operatorname{det} M}\\
		\alpha _{0} \cos (\theta)+w \sin (\theta)\\
		-\alpha _{0} \sin (\theta)+w \cos (\theta)\\
		q
	\end{bmatrix}
	\label{sys}
\end{align}
where $w$ is the heave velocity, $q$ is the pitch velocity, $x$ and $z$ are the horizontal and vertical positions respectively, and $\theta$ is the pitch angle. The cruise speed $ \alpha _0 $ is set to 4.11 m/s. The mass matrix elements include: $ m_{11} = m - Z_{\dot w} $ denotes the effective mass, $ m_{12} = -Z_{\dot q} $ represents the effect of the pitching speed on the transverse force, $ m_{22} = I_y - M_{\dot q} $ denotes the effective inertia for the pitching motion, and $ m _{21} = -M_{\dot w} $ is the component of the effect of ascent velocity on the pitching moment, $ \text{detM} $ is the determinant of the mass matrix.
\begin{align}
	Z = Z_q \cdot q + Z_w \cdot w + Z_\delta \cdot \delta
\end{align}
\begin{align}
	M = M_q \cdot q + M_w \cdot w + M_\theta \cdot \theta + M_\delta \cdot \delta
\end{align}
where $ Z_q $, $ Z_w $ and $ Z_\delta $ are the force components of the pitch velocity $q$, the ascent velocity $w $ and the system input rudder angle $\delta $, respectively. $ M_q $, $ M_w $, $ M_\theta $ and $ M_\delta $ are the moment components of the state variables. We give the following nonlinear data for the DSRV:
\begin{table}[h]  
	\centering  
	\caption{Values of DSRV variables}  
	\begin{tabular}{p{3.8cm} p{2cm}} 
		\hline  
		\textbf{Parameter} & \textbf{Value} \\ \hline  
		\ \ \ \ \	$U_0$           & $4.11$           \\  
		\ \ \ \ \	$m_{11}$        & $0.067936$       \\  
		\ \ \ \ \	$m_{12}$        & $0.000130$       \\  
		\ \ \ \ \	$m_{21}$        & $0.000146$       \\  
		\ \ \ \ \	$m_{22}$        & $0.003498$       \\  
		\ \ \ \ \	$Z_q$           & $-0.017455$      \\  
		\ \ \ \ \	$Z_w$           & $-0.043938$      \\  
		\ \ \ \ \	$Z_\delta$      & $0.027695$       \\  
		\ \ \ \ \	$M_q$           & $-0.01131$       \\  
		\ \ \ \ \	$M_w$           & $0.011175$       \\  
		\ \ \ \ \	$M_\theta$      & $-0.156276 / \alpha ^2$\\  
		\ \ \ \ \	$M_\delta$      & $-0.012797$      \\ \hline  
	\end{tabular}  
	\label{tab:dsrv_parameters_values}  
\end{table}

The DSRV is designed with a length of $L=5$ m. The control objective is to dive to a target depth of 50 m. The control input $u$, which corresponds to the rudder angle adjustment, is constrained within the range $[-30^\circ, 30^\circ]$. In the simulation, 700 $Z^n$ nodes and 400 $H^m$ nodes are used. The controller adjusts the diving speed and the attitude of the submersible by varying the rudder angle to ensure accurate control towards the target depth.

Data collection is carried out by simulating multiple random initial conditions over $500$ trajectories, each with $300$ time steps. The dynamics are discretized using a 4th-order Runge-Kutta method with a step size of $\Delta t = 0.01$ s. The collected data is used to train a BLS, where the system states are lifted into a higher-dimensional space using $700$ feature extraction nodes and $400$ enhancement nodes, resulting in a total of $1105$ lifted dimensions. The upscaled state is generated by the following equation:
\begin{align}
	\Phi _B(\xi)=\begin{bmatrix}
		\xi &Z_1(\xi)   & \dots  &Z_{700}(\xi)  &H_1(\xi)    &\dots   &H_{400}(\xi)
	\end{bmatrix}
\end{align}
\begin{align}
	\Phi _B(y)=\begin{bmatrix}
		y &Z_1(y)  & \dots  &Z_{700}(y)  &H_1(y)    &\dots   &H_{400}(y)
	\end{bmatrix}
\end{align}
Where $z_i$ and $H_j$ sub-tables represent feature nodes and enhancement nodes. After obtaining the upscaled state representation, the matrix $A_b$, $B_b$ and $C_b$ is solved by $(\ref{A})-(\ref{Cb})$ and the high-dimensional predictor (\ref{yuce}) is obtained.
\begin{figure}[!t]  
	\centering  
	\subfloat{\includegraphics[width=3.1in]{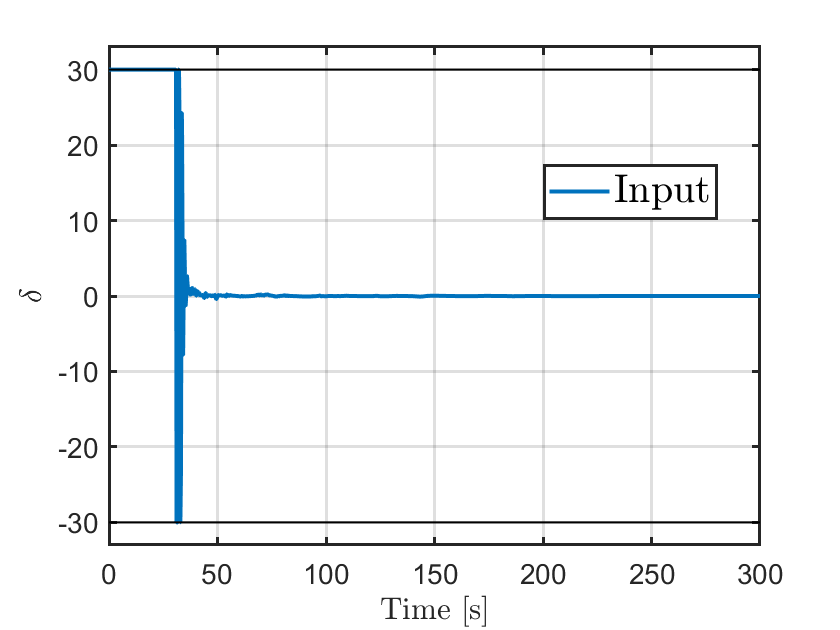}%
		\label{fig1}}  
	\hfill  
	\subfloat{\includegraphics[width=3.1in]{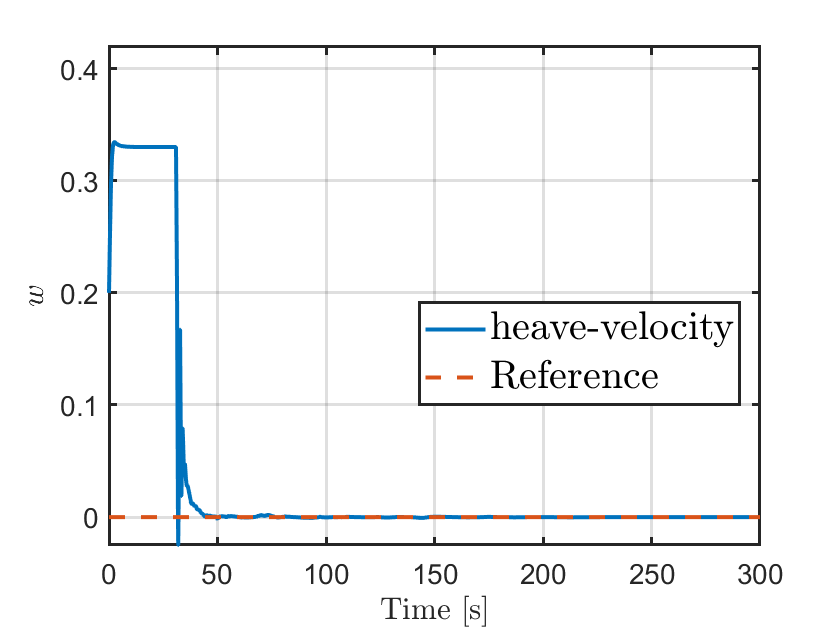}%
		\label{fig2}} 
	\hfill  
	\subfloat{\includegraphics[width=3.1in]{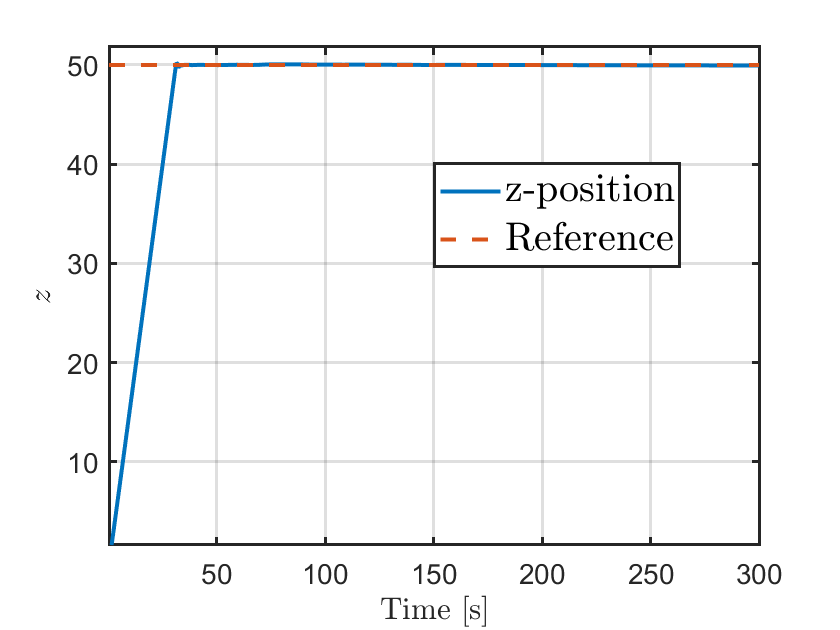}%
		\label{fig3}}  
	
	\caption{BE-MPC based UUV dive depth $50$ m position task: rudder angle $[-30^\circ, 30^\circ]$.} 
	\label{MPC}  
\end{figure}
\begin{remark}
	We design the model predictive controller by selecting only $w$ and $z$ in the state as state outputs, so $C_b=\begin{bmatrix} 1 & 0 & 0 & 0 & 0 & \cdots & 0 & 0 \\ 0 & 0 & 0 & 1 & 0 & \cdots & 0 & 0 \end{bmatrix}_{2\times1100}$, and $Q\in \mathbb{R}^{2\times2}$ simplifies the problem and focuses on the key states. At the same time, the computational cost of this method is comparable to that of standard linear MPC over the same prediction time domain.
\end{remark}

Based on the predictor in (\ref{yuce}), design the objective function (\ref{youhua}), The simulation lasts $300$ s and the prediction step $N_t$ is set to $20$. Since we are concerned with the state of motion in the vertical direction, we set $Q=\begin{bmatrix}
	10& 0\\
	0& 50\\
	
\end{bmatrix}$, $R=0.1$. As shown in Fig. \ref{MPC}, the simulation results demonstrate that the BE-MPC algorithm successfully achieves precise control of the DSRV at the target depth of $z = 50$ m. The figure illustrates the dynamic changes in the submersible's diving depth, vertical velocity, and control inputs (rudder angle) throughout the simulation. The initial state was selected as $x_0= [0.2; 0; 0; 0.1; 0]^T$. The control input (rudder angle) was constrained within the range of $[-30°, 30°]$. From the figure, it can be observed that the controller effectively adjusts the rudder angle to optimize the motion trajectory and attitude control. During the entire simulation process, the vertical position of the submersible steadily converges to the target depth of 50 meters. This verifies that the BE-MPC algorithm can successfully perform precise control tasks in a complex nonlinear dynamic system.

\section{Conclusions}
In this paper, we propose a BLS-EDMD method for approximating the Koopman operator and use this method to design a new MPC. This method solves the problem of model prediction error caused by the randomness associated with the choice of basis functions in the traditional EDMD method. By leveraging the feature and enhancement layers of the BLS network, the BLS-EDMD method refines the generation of basis functions, thereby enhancing the system state representation and boosting the model's overall prediction accuracy. In simulation experiments, we apply the proposed method to the target tracking control task of a classical van der Pol oscillator system and a DSRV. Among the prediction experiments, the results show that the BLS-EDMD-based Koopman predictor has significant advantages in terms of accuracy and stability. In addition, the control accuracy of the BLS-EDMD-based MPC controller is effectively validated in a complex DSRV scenario, highlighting its potential to effectively handle high-dimensional nonlinear systems.

In future work we will further extend the current BE-MPC methodology for applications in more complex nonlinear dynamic systems, such as multi-degree-of-freedom UUV and underwater devices with more complex hydrodynamic properties.


\vfill

\end{document}